\documentclass[prd,nofootinbib,preprintnumbers,a4paper]{revtex4}
\usepackage[a4paper,left=1in,right=1in,top=1.4in,bottom=1.4in]{geometry}
\usepackage{amsmath}
\usepackage{amssymb}
\usepackage{graphicx}
\usepackage{color}
\usepackage{xspace}
\usepackage{hyperref}

\newcommand{\GeV}{\ensuremath{\,\mathrm{GeV}}\xspace}
\newcommand{\TeV}{\ensuremath{\,\mathrm{TeV}}\xspace}
\newcommand{\jet}{\ensuremath{\,\mathrm{jet}}\xspace}
\newcommand{\as}{\ensuremath{\alpha_s}}
\newcommand{\aEW}{\ensuremath{\alpha_\text{EW}}}
\newcommand{\nLO}{\nbar \text{LO}\xspace}
\newcommand{\nNLO}{\nbar \text{NLO}\xspace}
\newcommand{\LO}{\text{LO}\xspace}
\newcommand{\NLO}{\text{NLO}\xspace}
\newcommand{\NNLO}{\text{NNLO}\xspace}
\newcommand{\nbar}{\ensuremath{\bar{n}}}
\newcommand{\order}[1]{\mathcal{O}\!\left(#1\right)}
\newcommand{\eg}{e.g.}
\newcommand{\ie}{i.e.}
\newcommand{\ETmissproj}{\ensuremath\text{\it projected } E_{T,\text{miss}}}
\newcommand{\LS}{{\ensuremath{\,\mathrm{LS}}\xspace}}

\begin{document}

\title{W\boldmath{$^{+}$}W\boldmath{$^{-}$} production at high transverse
momenta beyond NLO}

\author{Francisco Campanario}
\email{francisco.campanario@ific.uv.es}
\affiliation{Theory Division, IFIC, University of Valencia-CSIC,
E-46980 Paterna, Valencia, Spain}
\author{Michael Rauch}
\email{michael.rauch@kit.edu}
\affiliation{Institute for Theoretical Physics, Karlsruhe Institute of
Technology (KIT), Germany}
\author{Sebastian Sapeta}
\email{sebastian.sapeta@durham.ac.uk}
\affiliation{Institute for Particle Physics Phenomenology, Durham University,
South Rd, Durham DH1 3LE, United Kingdom}

\preprint{DCPT/13/146}
\preprint{IFIC/13-68}
\preprint{IPPP/13/73}
\preprint{FTUV-13-0927}
\preprint{KA-TP-24-2013}
\preprint{LPN13-064}
\preprint{SFB/CPP-13-66}

%%%%%%%%%%%%%%%%
\begin{abstract}
\vskip 10pt
Pair production of W gauge bosons is an important process at the LHC  entering
many experimental analyses, both as background in new-physics searches or 
Higgs measurements and as signal in precision studies and tests of the Standard
Model. Therefore, accurate predictions for this class of processes are of great
interest in order to exploit the full potential of LHC measurements.
We use the LoopSim method to combine NLO QCD results for WW and
WW+jet, as well as the loop-squared gluon-fusion contribution, to obtain
approximate NNLO predictions for WW production. The cross sections are
calculated with VBFNLO and include leptonic decays of the W bosons as well as
finite-width and off-shell effects. 
We find that the size of the additional corrections beyond NLO can be
significant and well outside of the NLO error bands given by renormalization and
factorization scale variation. Applying a jet veto, we observe further negative
corrections at NNLO, which we relate to the presence of large Sudakov
logarithms.

\end{abstract}
%%%%%%%%%%%%%%%%

\maketitle

%-----------------------------------------------------------------------------
\section{Introduction}

The production of two electroweak bosons constitutes an important process at
the Large Hadron Collider~(LHC), in particular when combined with the
leptonic decay modes of the vector bosons. Of particular interest in
this class of processes is the production of two opposite-sign W bosons,
which we will consider in this article.

When letting the W bosons decay leptonically, this process has two
charged leptons of opposite sign and either same or different flavour,
and two neutrinos in the final state
\begin{equation}
pp \rightarrow W^+W^- + X \rightarrow \ell_1^+ \nu_1 \ell_2^- \bar{\nu}_2 +X \ .
\end{equation} 
Due to the presence of neutrinos, it is not possible to reconstruct the
invariant mass of each individual W boson. Therefore, this
process is an important background for many measurements at the LHC,
for example the searches and studies of Higgs
bosons, where it forms an irreducible background to the WW decay mode
of the Higgs~\cite{ATLASHiggsWW, CMSHiggsWW}. 
It is also an important background process to searches for physics beyond the
Standard Model (BSM), which often contains a light stable particle that
manifests itself as missing transverse energy in the
detector~\cite{Feigl:2012df,Curtin:2012nn,Aad:2012nev}.  But the WW production
is also interesting in its own right, as it allows to test the SM, \eg{} when
investigating anomalous triple gauge couplings (aTGC), where it allows to put
stringent constraints on parameters like $\Delta
g_1^Z$~\cite{ATLAS:2012mec,Chatrchyan:2013yaa,Chatrchyan:2013oev}.

From the experimental side, WW production has been measured at the LHC in the
7~TeV run by both ATLAS and CMS experiments~\cite{ATLAS:2012mec,Aad:2012oea,
Chatrchyan:2011tz,Chatrchyan:2013yaa}, and from CMS also first results at 8~TeV,
from the 2012 run, with approximately~5~fb$^{-1}$ of integrated luminosity, are
available~\cite{Chatrchyan:2013oev}. These results are in reasonable agreement
with the SM predictions, with both experiments observing slightly more events
than expected but still within two standard deviations range.

On the theory side, the next-to-leading order~(NLO) QCD
corrections for WW production have been computed in
Ref.~\cite{dibosonQCD}. These corrections turn out to be large, of the
order of 50\% at the level of the total cross section for inclusive
cuts. This is mainly due to new contributions appearing in
the real-emission part, namely gluon-initiated channels. The large
size of the gluon parton density functions (PDFs) partly compensates the
suppression from the additional power of the strong coupling constant
$\alpha_s$. Looking at differential distributions, their K-factors,
defined as the ratios at NLO over LO, are also sizable and very often
phase-space dependent. This means that a simple multiplication of the LO
distributions with the K-factor of the integrated result (which is just a single
number) will not reproduce the full NLO distribution correctly. Thus,
flexible and fast NLO Monte-Carlo programs are needed to obtain reliable
predictions.
Adding soft-gluon resummation of threshold logarithms gives a mild
enhancement of the cross section~\cite{Dawson:2013lya}.
Work on the full NNLO QCD corrections to WW
production process has only been started~\cite{Gehrmann:2013cxs}.

Gluon-initiated contributions~\cite{gginduced}, with a closed fermion loop,
formally appear only at NNLO QCD. They cannot interfere with the tree-level
diagrams and therefore enter as one-loop squared diagrams. However, due to the
large gluon PDFs, their numerical impact is relevant, giving a contribution at
the 10\% level in typical Higgs analyses and at the 3-5$\%$ level in inclusive
searches. 

Electro-weak corrections have been reported for on-shell production in
Refs.~\cite{dibosonEW}. While their effect is usually small for
integrated cross sections, the tails of differential distributions can
receive sizable corrections in typical experimental analysis setups. 

Results for WWj at NLO QCD, which provide one-loop real-virtual
and double-real contributions to the NNLO corrections for WW production,
have been computed in a series of papers~\cite{dibosjet}. The integrated
corrections to the LO result are typically in the 40\% range, while again the
tails of differential distributions can show significantly larger K-factors.

NLO QCD corrections to the double real-emission process WWjj are also
available~\cite{Greiner:2012im}. Here, no new processes open up at
higher-order and hence the integrated NLO result shifts by a
moderate value of about 10\% relative to LO, with a greatly reduced scale
uncertainty.

Because the NLO QCD correction to WW production turns out to be large, it is
important to assess the size of the NNLO QCD corrections. WWj at NLO QCD
provides an essential piece of the WW at NNLO QCD result, accounting for new
sub-processes and new topologies appearing for the first time at NNLO. However,
this alone is not enough as it misses the 2-loop contributions which are needed
to cancel divergencies of the double-real and real-virtual diagrams.
Therefore, we use the LoopSim approach~\cite{Rubin:2010xp} to simulate 
2-loop contributions to the WW process and combine them with the tree and 
1-loop parts from WW and WWj at NLO.
As explained in next section, this gives us a dominant part of the NNLO
correction for a number of relevant observables.
 
Very recently, a related study of WW production with 0 and 1-jets in the final
state, using a different framework, has been presented in
Ref.~\cite{Cascioli:2013gfa}.

The outline of the paper is as follows. In section~\ref{sec:framework} we 
shortly recap the theoretical framework of combining NLO calculations.  Then, in
section~\ref{sec:setup}, we define the used model parameters and cuts, which
closely follow the experimental analyses. Finally, in section~\ref{sec:results}
we present the results of our calculation both for integrated cross sections as
well as several important distributions.
We close the paper with a brief summary of our findings in
section~\ref{sec:concl}.

%-----------------------------------------------------------------------------
\section{Theoretical Framework}
\label{sec:framework}

To compute approximate NNLO results for WW production, we use the LoopSim
approach~\cite{Rubin:2010xp} to merge WW@NLO and WWj@NLO samples obtained with
the VBFNLO package~\cite{VBFNLO}.
The WWj@NLO result, provided by VBFNLO, constitutes the
double-real and real-virtual parts of WW@NNLO. These parts alone are,
however, not sufficient as they diverge upon integration over the phase
space of the real partons.
Those divergences are bound to be canceled, for sufficiently inclusive
observables, by the 2-loop virtual correction, following the
KLN~\cite{KLN} theorem.
Due to this cancellation, it is possible to construct the dominant part
of those 2-loop diagrams from their corresponding real emission
counterpart using the LoopSim method.  Thereby, the singular structure
matches exactly the one from real diagrams with higher multiplicity.
 
LoopSim is based on unitarity and starts from assigning an approximate
angular-ordered branching structure to each WWj@NLO event with the help of the
Cambridge/Aachen (C/A)~\cite{Dokshitzer:1997in, Wobisch:1998wt} jet algorithm,
with a given radius $R_\text{LS}$. 
Then, the underlying hard structure of the event is determined and the
corresponding particles are marked as ``Born''. The number of Born
particles is fixed for a given process, and it is given by the number of
final state particles at tree level. For simplicity, we combine the
neutrino and anti-lepton and construct a virtual W$^+$ with the
four-momentum given by the sum of the two daughter particles, and
similarly for W$^-$.  Therefore, the number of Born particles for WW
production is 2. At NNLO, these can be either two vector bosons, a
boson and a parton, or two partons. 
 
The remaining particles, which were not marked as ``Born'', are then
``looped'' by finding all possible ways of recombining them with the
emitters. 
In this step, LoopSim generates an approximate set of 1 and 2-loop
diagrams with weights equal to $(-1)^\text{number of loops}$ times the
weight of the original event. Finally, a double counting between the
approximate 1-loop events generated by LoopSim and the exact 1-loop
events from the WWj@NLO sample is removed. 
To distinguish our results, with simulated loops, from the exact ones, we denote
the approximate loops by $\nbar$, as opposed to N used for exact loops. So, for
example, $\nLO$ means the correction with simulated 1-loop diagrams, but $\nNLO$
is a result with exact 1-loop diagrams and simulated 2-loop contributions.

Note that, because the weights of the loop diagrams produced by LoopSim come
from the corresponding real diagrams with higher multiplicity and differ only by
the sign, the sum of weights for a set of events generated by LoopSim from a given
original event must be zero (see~\cite{Rubin:2010xp} for a detailed explanation).
The latter follows from unitarity and means that, for the fully inclusive case,
the \nNLO integrated cross section is equal to the NLO one. 
However, because the W bosons are not necessarily identified as ``Born''
particles, LoopSim will use this type of events to generate new diagrams with
simulated W loops that will not contribute to our result, as we require two W
bosons in the final state. Similarly, other cuts imposed on leptons, missing
energy and jets will spoil the exact real-virtual cancellation outside of the
soft or collinear limit, leading to a difference between \nNLO and NLO
integrated cross sections.

Our approximate \nNLO results will have exact tree and 1-loop parts, and the
exact singular part of the 2-loop diagrams. Hence, the results will be
finite and they will differ from the full NNLO only by the constant
terms of the 2-loop contribution.
This difference is, however, very small for an observable, $A$, that receives
significant NLO corrections due to new channels or new topologies
\begin{equation}
  \frac{d\sigma^\text{\nNLO}}{dA} - \frac{d\sigma^\text{\NNLO}}{dA} = 
  {\cal O}\left(\alpha_s^2\, \frac{d\sigma^\text{\LO}}{dA}\right)\,.
  \label{eq:ls-accuracy}
\end{equation}
Therefore, for this type of observables, our \nNLO result will contain the
dominant part of the NNLO prediction for WW production. 

One class of uncertainties of the LoopSim method is probed by varying the
$R_\text{LS}$ parameter. It accounts for the part of the procedure 
related to attributing the emission sequence and the underlying hard structure
of the events~\cite{Rubin:2010xp}.
The smaller the value of $R_\LS$, the more likely the
particles are recombined with the beam, the larger $R_\LS$, the more likely they
are recombined together. 
The value of $R_\text{LS}$ affects therefore only the wide-angle or hard
emissions where the $ij$ mergings compete with the mergings with the beam.
In this study, we use $R_\LS = 1$ and vary it by $\pm 0.5$.  As
we shall see in section~\ref{sec:results}, the uncertainty related to
$R_\text{LS}$ is smaller than that coming from renormalization and factorization
scale variation, except for the very low $p_T$ region.

In order to make the communication between VBFNLO and LoopSim possible, an
interface has been created~\cite{Campanario:2012fk}, which is based on the Les
Houches Event (LHE) format~\cite{Alwall:2006yp} used to pass the information
between the two programs.

%%%%%%%%%%%%%%%%%%%%%%%%%%%%%%%%%%%%%%%%%%%%%%%%%%%%%%%%%%%%%%%%%%%%%%%%%%%%%%
\section{Calculational setup}
\label{sec:setup}

In our calculation, we take as input parameters in the electroweak
sector the masses of the W, Z and Higgs boson, and the Fermi constant.
The electromagnetic coupling constant and the weak mixing angle are then
derived from the above via electro-weak tree-level relations:
\begin{align}
m_Z &= 91.1876 \GeV\,, & G_F &= 1.16637 \times 10^{-5}\GeV^{-2}\,, \nonumber\\
m_W &= 80.398  \GeV\,, & \alpha_\text{em}^{-1} &= 132.3407     \,, \nonumber\\
m_H &= 126     \GeV\,, & \sin^2(\theta_W) &= 0.22265\ .
\end{align}
Finite-width effects in propagators with massive gauge
bosons are taken into account using a modified
version~\cite{Alwall:2007st,Oleari:2003tc} of the
complex-mass scheme~\cite{Denner:1999gp}, where $\sin^2(\theta_W)$ is
kept real. As numerical values, we use $\Gamma_W = 2.097\GeV$ and
$\Gamma_Z = 2.508\GeV$.
The effects of external top or bottom quarks are neglected, but their
contribution is taken into account in the closed fermion loops appearing
in the gluon-fusion part. We use the following values of top and bottom masses
\begin{align}
m_t &= 172.4 \GeV\,, & m_b = m_b(m_H) = 2.84 \GeV \ .
\end{align}
We choose the running bottom mass at the Higgs mass as its
largest contribution is in the mediation of the effective ggH coupling,
where this choice turns out to be advantageous. 
Other choices like an on-shell mass lead to cross section changes at the
sub-percent level if considering the continuum box part alone, and
are a few percent for the full (box+Higgs) gluon-fusion contribution.
Hence, the difference is at the per mill level for the cross section of
the full process and the exact choice does not play any role.
All other quarks are treated as massless and any quark mixing effects are
neglected.
Regardless of the order, we take the MSTW NNLO 2008~\cite{Martin:2009iq}
PDF set with $\alpha_s(m_Z)= 0.11707$, using the implementation provided
by LHAPDF~\cite{LHAPDF}.
The final state partons, if any, are clustered with the anti-$k_t$
algorithm~\cite{Cacciari:2008gp}, with radius $R=0.5$, as implemented in
FastJet~\cite{Cacciari:2005hq, Cacciari:2011ma}.

As the central value for the factorization and renormalization scales we choose
\begin{equation}
\mu_{F,R}= \frac12 \left( 
\sum  p_{T,\text{partons}} + 
\sqrt{p_{T,W^+}^2+m_{W^+}^2}+
\sqrt{p_{T,W^-}^2+m_{W^-}^2} 
\right) \,,
\label{eq:ren}
\end{equation}
where $p_{T,W^{\pm}}$ and $m_{W^\pm}$ are the transverse momenta and
invariant masses of the decaying W bosons, respectively.

In our phenomenological analysis presented in the next section, we
include both electron and muon decay channels of the W boson, in both
same-flavour, $e^+e^-$ and $\mu^+\mu^-$, and different-flavour, $e^\pm
\mu^\mp$ variants.
Our fiducial volume matches to a large extent that chosen by the
CMS experiment in Ref.~\cite{Chatrchyan:2013oev}. 
The ATLAS setup is also very similar, although the exact numerical values of the
cuts differ~\cite{ATLAS:2012mec,Aad:2012oea}.
 
All events are required to have a pair of oppositely charged leptons 
of either the first or the second generation (same-flavour case) or one
from the first and one from the second generation (different-flavour).
Both of them must fulfill the cuts 
\begin{equation}
p_{T,\ell} > 20 \GeV \qquad \text{ and } \qquad |\eta_{\ell}| < 2.5 \ .
\end{equation}

A $\ETmissproj$ is defined, following Ref.~\cite{Chatrchyan:2013oev}, as the
missing transverse energy, if the angle between the missing transverse momentum
and the lepton closest in azimuthal angle is larger than $\pi/2$, or its
component transverse to the closest lepton direction
otherwise.  For different-flavour configurations, we require that the
$\ETmissproj > 20 \GeV$.

In the same-flavour case, we use a more restrictive cut with $\ETmissproj
> 45 \GeV$. Moreover, we select only events
with dilepton mass $m_{\ell\ell} > 12 \GeV$, $|m_{\ell\ell}- m_Z|
> 15 \GeV$ and dilepton transverse momentum $p_{T,\ell\ell} > 45 \GeV$.
For these same-flavour runs, we also require that the angle in the transverse
plane between the dilepton system and the most energetic jet with $p_T> 15 \GeV$
is smaller than 165 degrees.
These additional dilepton cuts are used by the experiments to reduce Drell-Yan background and jets misidentified as leptons.

In our study, we shall discuss two classes of results -- without and with
jet veto. For the latter case, we reject all events containing one or more
jets with $p_{T,\jet} > 30 \GeV$ and $|\eta_{\jet}| < 4.7$. These vetoed
results are particularly important when discussing the impact of our
findings on the experimental results. The measurements of the inclusive
WW cross section by both ATLAS and CMS have been performed with a vetoed
setup~\cite{Aad:2012oea,ATLAS:2012mec,Chatrchyan:2013yaa,Chatrchyan:2013oev,
Chatrchyan:2011tz}. Also in the studies and searches of the Higgs boson, events
are grouped into categories with different jet multiplicity. For the lowest
0-jet bin this effectively corresponds to a jet veto, and for higher ones
correspondingly to n-jet exclusive samples with vetoes on additional jets
beyond the desired number~\cite{ATLASHiggsWW,CMSHiggsWW}. 
As we shall see in the following section, these two classes
of results exhibit distinctly different behaviour of higher order
perturbative QCD corrections.

%-----------------------------------------------------------------------------
\section{Numerical Results}
\label{sec:results}

%%%
\begin{table}
\scalebox{1.2}{
\begin{tabular}{l|rrll|rrll}
& \multicolumn{4}{|c|}{c.s. [fb] without jet veto} 
& \multicolumn{3}{|c}{c.s. [fb] with jet veto} \\\hline 
$\sigma_\text{LO}$          &   247.49 & ${}^{+5.40}_{-7.60}$ & & 
                            &   247.49 & ${}^{+5.40}_{-7.60}$ &
                            \\[1ex]
$\sigma_\text{box+Higgs}$   &   19.02 & ${}^{-3.70}_{+4.86}$ & & 
                            &   19.02 & ${}^{-3.70}_{+4.86}$ &
                            \\[1ex]
$\sigma_\text{pure-NLO}$    &  334.64 & ${}^{-6.36}_{+6.49}$ & & 
                            &  253.05 & ${}^{+2.98}_{-4.75}$ &
                            \\[1ex]
$\sigma_\text{pure-\nNLO}$  &  345.17 & ${}^{-7.06}_{+7.03}$ 
                            &   ($\mu$) ${}^{+5.24}_{-3.33}$ ($R_\text{LS}$)& 
                            &  236.63 & ${}^{-1.16}_{+1.45}$ 
                            &   ($\mu$) ${}^{+5.31}_{-3.27}$ ($R_\text{LS}$)&
                            \\[1ex]
\hline
$\sigma_\text{NLO}$         &  353.67 & ${}^{-10.06}_{+11.35}$ &  &
                            &  272.07 & ${}^{-8.45}_{+7.84}$ & &
                            \\[1ex]
$\sigma_\text{\nNLO}$       &  364.19 & ${}^{-10.76}_{+11.89}$ 
                            &   ($\mu$) ${}^{+5.24}_{-3.33} $ ($R_\text{LS}$) &
                            &  255.72 & ${}^{-4.86}_{+6.31}$ 
                            &   ($\mu$) ${}^{+5.31}_{-3.27} $ ($R_\text{LS}$) &
                            \\[1ex]
\end{tabular}
}
\caption{
Integrated (fiducial) cross sections for $pp \rightarrow W^+ W^- \rightarrow
\ell^+ \nu \ell^- \bar{\nu}$ at the LHC with $\sqrt{s} = 8 \TeV$ using the
parameter settings and cuts given in section~\ref{sec:setup}. 
According to the naming convention adopted throughout the paper, $\sigma_\NLO$
and $\sigma_\text{\nNLO}$, given in the last two rows, contain the gluon-fusion
contribution (labeled ``box+Higgs'' above). Pure NLO and \nNLO cross sections
(labeled with pure-NLO and pure-\nNLO) are given in the 3$^\text{rd}$ and the
4$^\text{th}$ row, respectively.
For the upper part of the table, the values in superscript correspond to the
renormalization and factorization scale $\mu = 2 \mu_0$, whereas those in
subscript refer to $\mu = \frac{1}{2} \mu_0$.
Similarly, for the \nNLO~results, we give the cross sections for the $R=1.5$ and
$R=0.5$ choices (each time with the central scale $\mu_0$) in the upper and
lower case, respectively.
The scale uncertainties of the NLO (\nNLO) cross sections, shown in the lower
part of the table, were obtained by linearly adding the respective positive and
negative uncertainties of the pure-NLO (pure-\nNLO) and the box+Higgs
contributions (see text for details).
The statistical error from Monte Carlo integration is at the per mill level for
all results.
}
\label{tab:cs}
\end{table}

In Table~\ref{tab:cs}, we present numerical results for integrated cross
sections. All results correspond to the LHC with a center-of-mass energy
$\sqrt{s} = 8 \TeV$ and are summed over same and different-flavour combinations in
the final state. 
We adopt the convention according to which $\sigma_\NLO$ and
$\sigma_\text{\nNLO}$, given in the last two rows of Table~\ref{tab:cs}, contain
also the gluon-fusion contribution (``box+Higgs''). Pure NLO and \nNLO cross
sections (labeled pure-NLO and pure-\nNLO) are given in the 3$^\text{rd}$
and the 4$^\text{rh}$ row of Table~\ref{tab:cs}, respectively.
The uncertainties are obtained by varying the renormalization and factorization
scale $\mu_R=\mu_F=\mu$ by the factors $2^{\pm 1}$. For the \nNLO cross
sections, we also give the errors from changing the radius parameter
$R_\text{LS}$ to $0.5$ and $1.5$.
The scale uncertainties of the NLO (\nNLO) cross sections were obtained by
linearly adding the uncertainties of the pure-NLO (pure-\nNLO) and the box+Higgs
contributions (see discussion below).

The central column of Table~\ref{tab:cs} shows the integrated cross sections without
imposing any cuts on additional jets.
Going from LO to pure-NLO, we observe corrections of the order of 35\%. 
The scale uncertainty is marginally reduced from a little bit above
2\% at LO to a little bit below 2\% at pure-NLO.
Merging the NLO results for WW and WWj with LoopSim gives the pure-\nNLO cross
section, which in the case without jet veto is about 3\% higher than pure-NLO.
The scale uncertainty is at the similar 2\% level. On top of that, the
uncertainty due to $R_\text{LS}$ variation is about 1.5\%. 

In the lower part of Table~\ref{tab:cs} we show the NLO and \nNLO cross
sections, which include the loop-squared gluon-fusion box and Higgs
contributions.  The scale uncertainties  were obtained by linearly adding the
respective positive and negative uncertainties of the pure-NLO (pure-\nNLO) and
the box+Higgs parts. Even though the box+Higgs contributes only about 5\% to the
NLO and \nNLO results, it comes with a relatively large scale uncertainty of
25\% of the box+Higgs cross section. This is because the gluon-fusion
contribution is effectively of leading order type as it enters for the first
time at $\order{\aEW^2\as^2}$.  Altogether, the scale uncertainties of
$\sigma_\NLO$ and $\sigma_\text{\nNLO}$ are at the level of 3\%.

\begin{figure}[t]
  \centering
  \includegraphics[width=0.49\columnwidth]{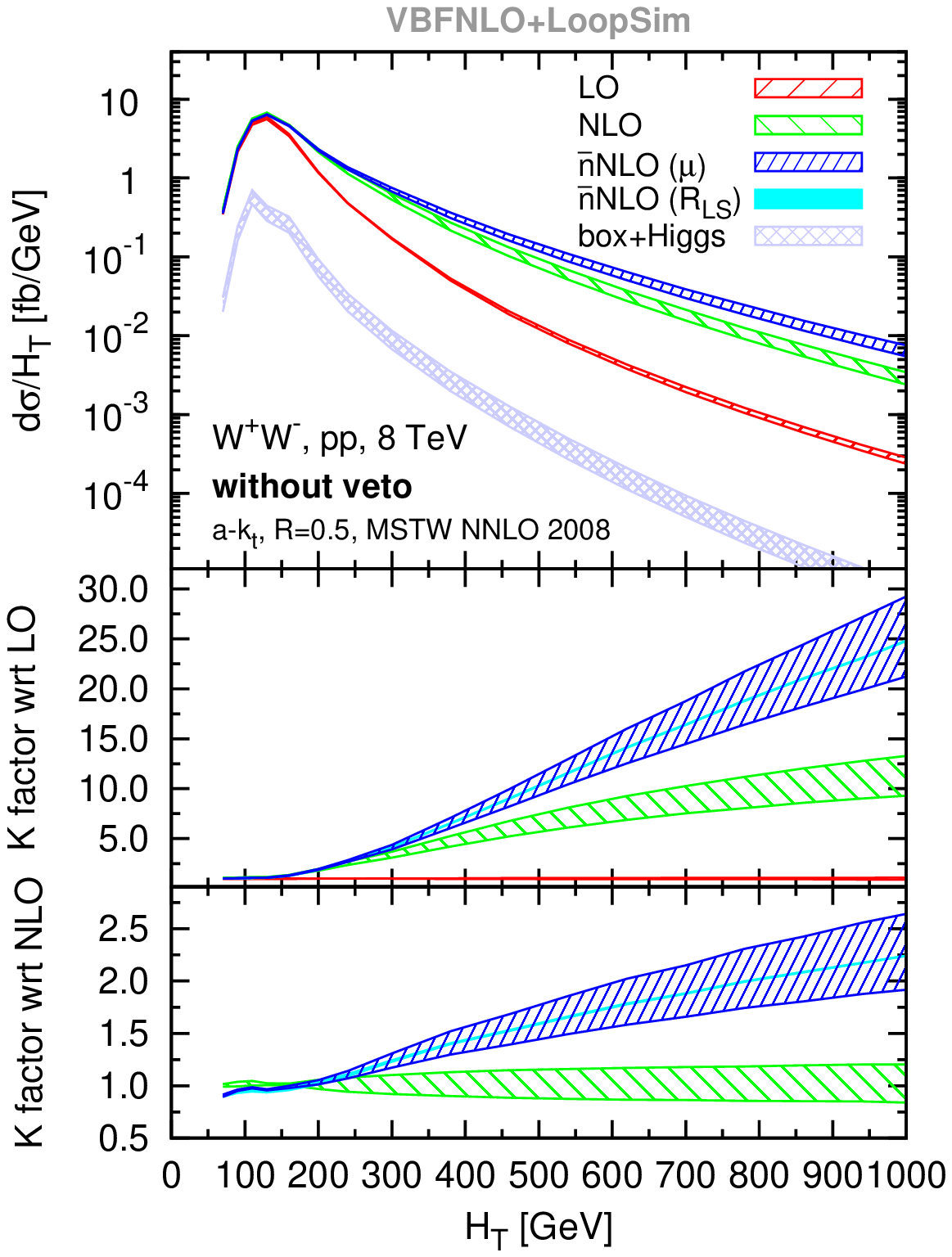}
  \hfill 
  \includegraphics[width=0.49\columnwidth]{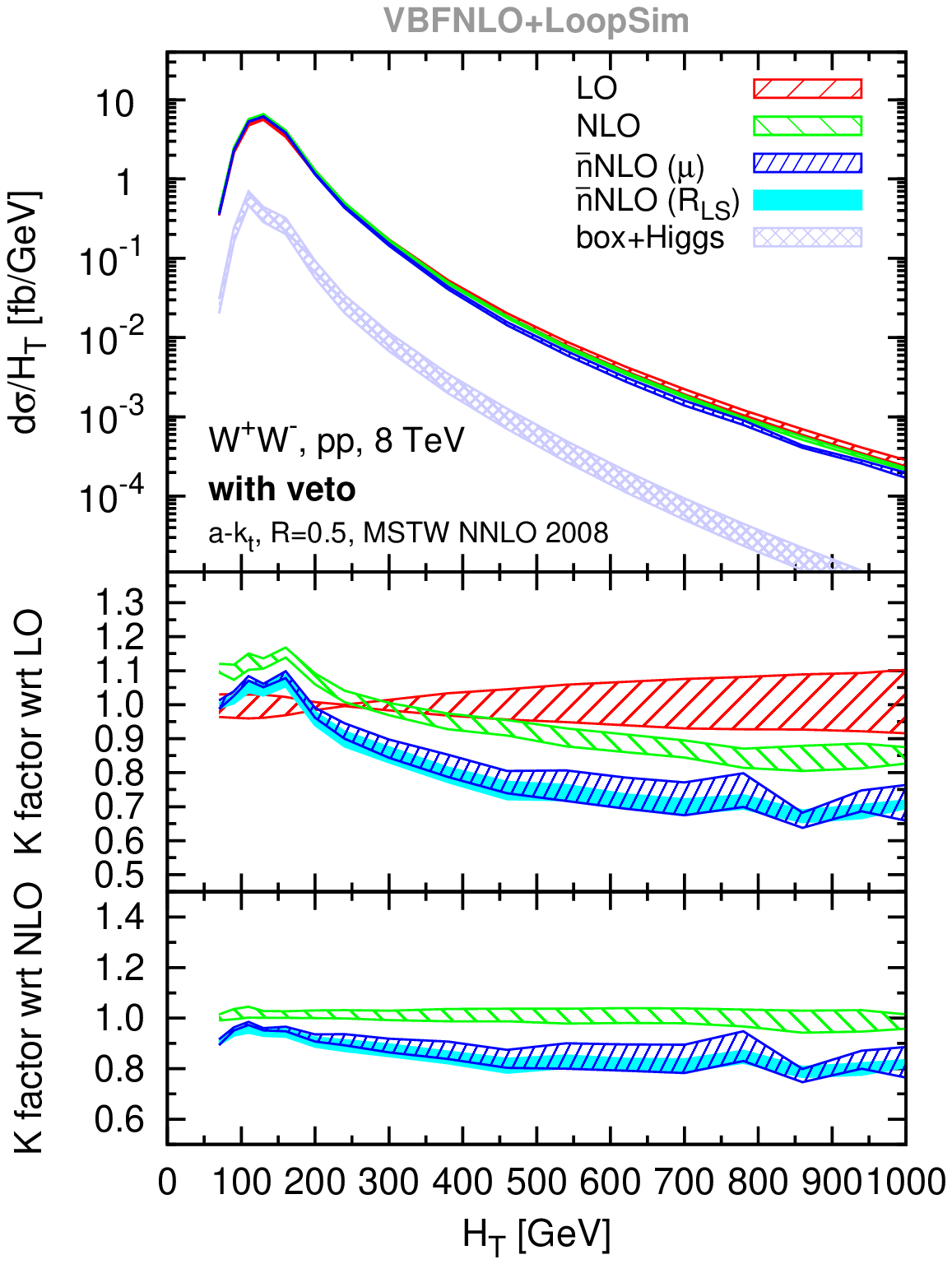}
  \caption{ %
  Differential cross sections and K-factors for the effective mass observable,
  defined in Eq.~(\ref{eq:HT}), for the LHC at $\sqrt{s}=8\, \text{TeV}$. The
  bands correspond to varying $\mu_F=\mu_R\equiv \mu$ by factors 1/2 and 2
  around the central value from Eq.~(\ref{eq:ren}). The cyan solid bands give
  the uncertainty related to the $R_\text{LS}$ parameter varied between 0.5 and
  1.5.
  The distribution is a sum of contributions from 
  $e^+  \nu_e   e^-  \bar{\nu}_e$,
  $\mu^+\nu_\mu \mu^-\bar{\nu}_\mu$, 
  $e^+  \nu_e   \mu^-\bar{\nu}_\mu$ and
  $\mu^+\nu_\mu e^-  \bar{\nu}_e$ decay channels.
  The contribution from the gluon-fusion box and Higgs diagrams is
  included in the NLO and \nNLO curves.
  The left panels correspond to the inclusive sample, while the results shown in
  the right panel were obtained with vetoing events containing jets
  which fulfill the criteria $p_{T,\jet} > 30 \GeV$ and $|\eta_{\jet}| < 4.7$.
  }
  \label{fig:HT}
\end{figure}

The right column of Table~\ref{tab:cs} shows the values of the cross sections
when a jet veto is imposed. The LO and box+Higgs results obviously do not differ
from the non-veto case as they have no partons in the final state. 
The K-factor from LO to pure-NLO is at the level of 2.5\%, much smaller than in
the case without veto. The scale uncertainty reduces by 40\%. Going one order
higher, it turns out that the $\order{\aEW^2\as^2}$ corrections are negative and
therefore the cross section decreases at pure-\nNLO. It also yields a reduced
scale uncertainty below 1\%. Hence, even if we combine it with the $R_\text{LS}$
error, the pure-\nNLO result in the vetoed case is still well below that at
pure-NLO.

We mentioned earlier that the uncertainties of the full NLO and \nNLO (hence
with box+Higgs) were obtained by linearly adding individual errors of each
contribution.
We choose this procedure since a naive scale variation of the sum of pure-NLO
and box+Higgs, for the case with jet veto, gives a nearly vanishing scale
uncertainty of $\sigma_\NLO$, which is a result of an accidental cancellation
between the pure NLO part and the gluon-fusion part.
That, in turn, is due to the fact that $\sigma_\text{pure-NLO}$ is an
increasing, while $\sigma_\text{box+Higgs}$ is a decreasing function of $\mu$.
Therefore they compensate each other in the sum, which changes only
by~${}_{+0.10}^{-0.72}$ upon scale variation.
We believe that adding errors of each individual contribution linearly gives a
much more realistic estimate of the uncertainty.
In all other cases, \ie{} pure-\nNLO with veto and pure-NLO and pure-\nNLO
without veto, the cross section decreases as function of $\mu$, just like in the
box+Higgs part. Hence, for those results, adding the errors linearly is
equivalent to the naive scale variation. Therefore, our procedure can be used
across all results given in Table~\ref{tab:cs}.

Finally, let us mention that the LoopSim method was designed to give 
an accurate estimate of the NNLO result predominantly at high-$p_T$, where the
constant term of 2-loop diagrams is of less importance. This constant term can
however bring significant contributions to the total cross section. 
Therefore, the \nNLO~result for the integrated (fiducial) cross section,
presented in Table~\ref{tab:cs}, captures the exact logarithmic terms of NNLO,
but only part of the constant terms, namely those coming from the tree and
1-loop diagrams. We believe that this adds additional information as 
it includes, for instance, the contributions from new partonic channels.
One should however still expect a genuine 2-loop correction on top of
the numbers given in Table~\ref{tab:cs}. 

%%%%%%%%%%%%%%%%%
\begin{figure}[t]
  \centering
  \includegraphics[width=0.49\columnwidth]{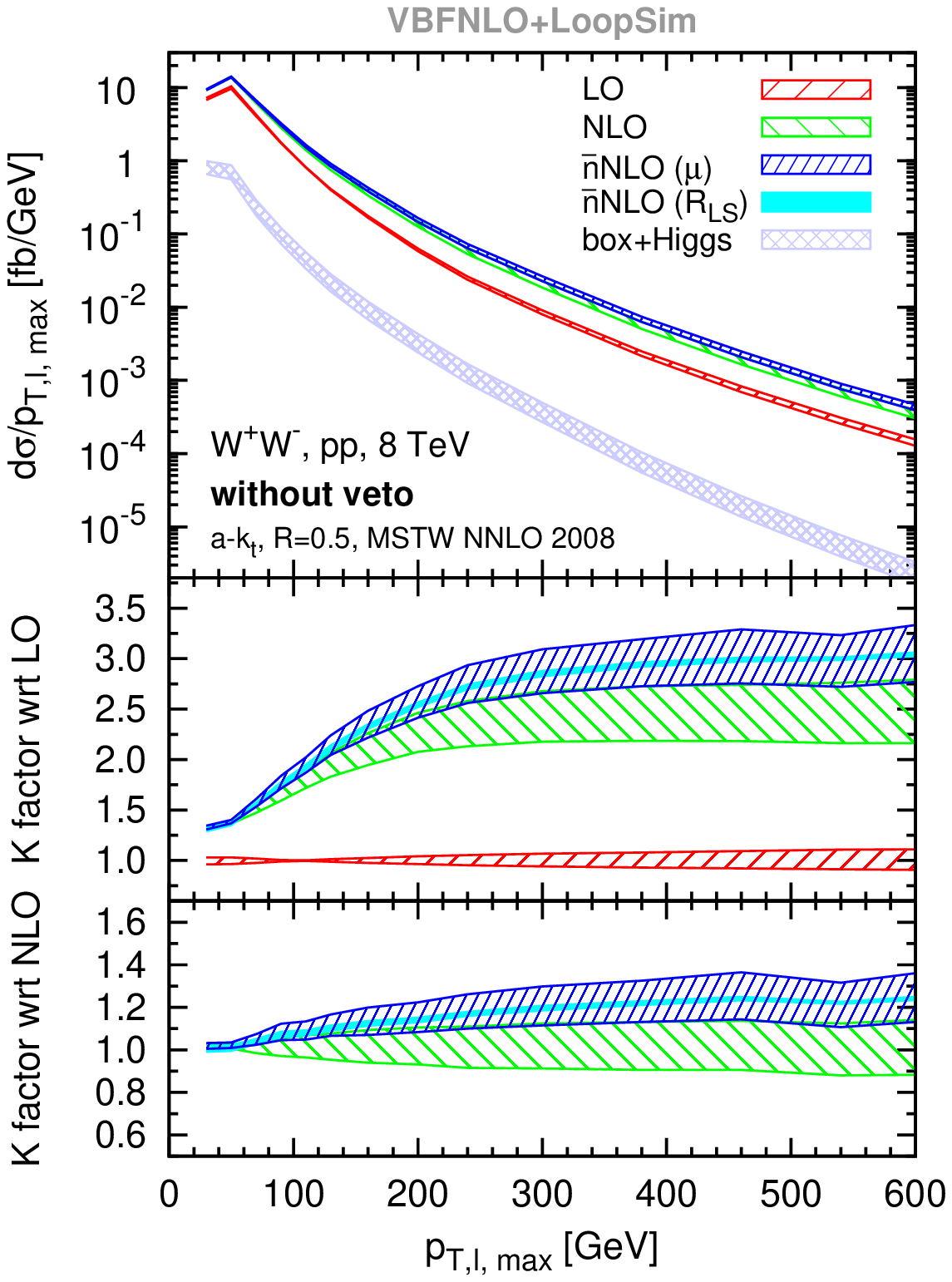}
  \hfill 
  \includegraphics[width=0.49\columnwidth]{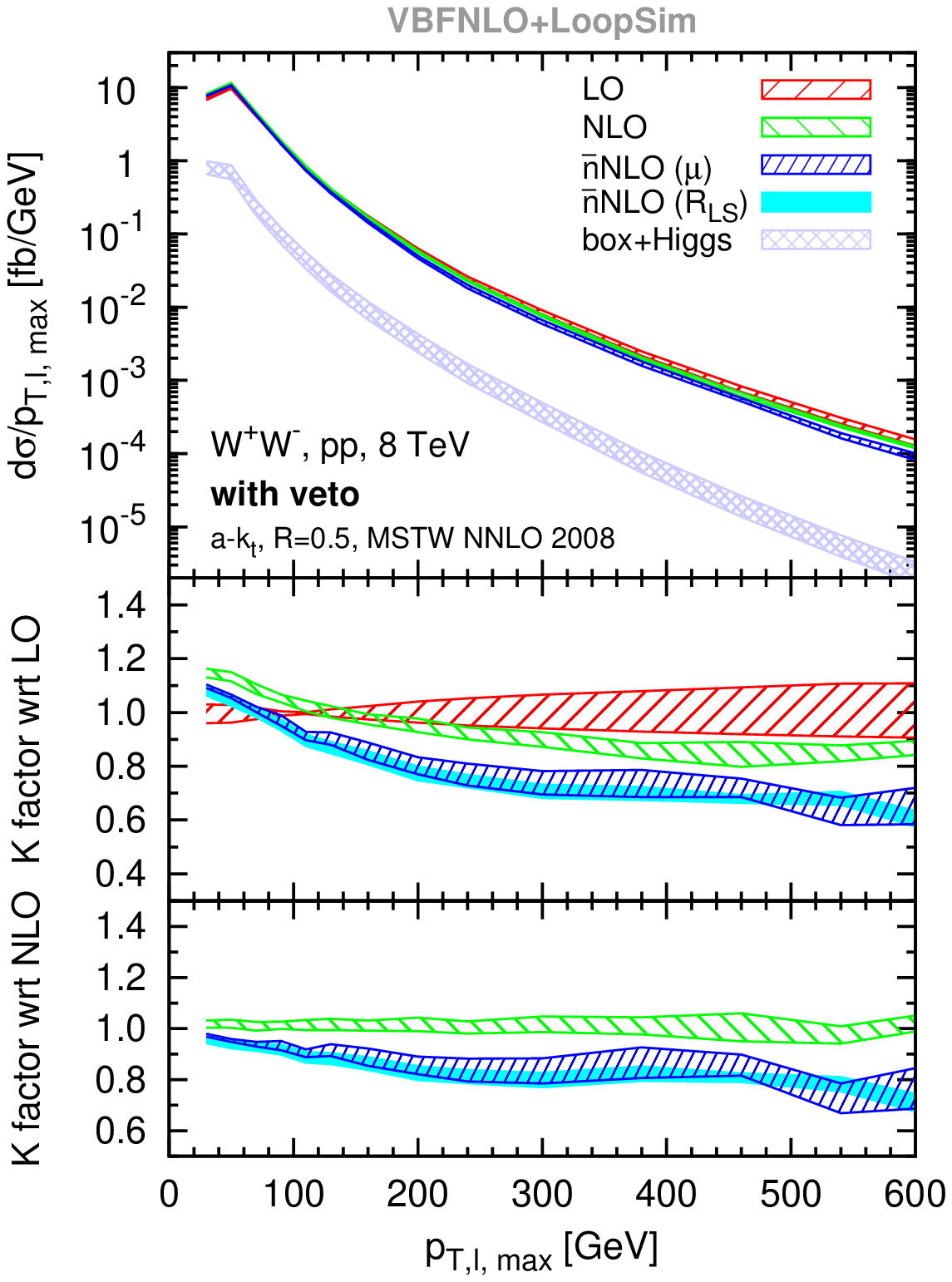}
  \caption{%
  Differential cross sections and K-factors for the $p_T$ of the hardest lepton
  for the LHC at $\sqrt{s}=8\, \text{TeV}$ without (left) and with
  jet veto (right). 
  Details are as in Fig.~\ref{fig:HT}.
  }
  \label{fig:ptlmax}
\end{figure}

%%%%%%%%%%%%%%%%%
\begin{figure}[t]
  \centering
  \includegraphics[width=0.49\columnwidth]{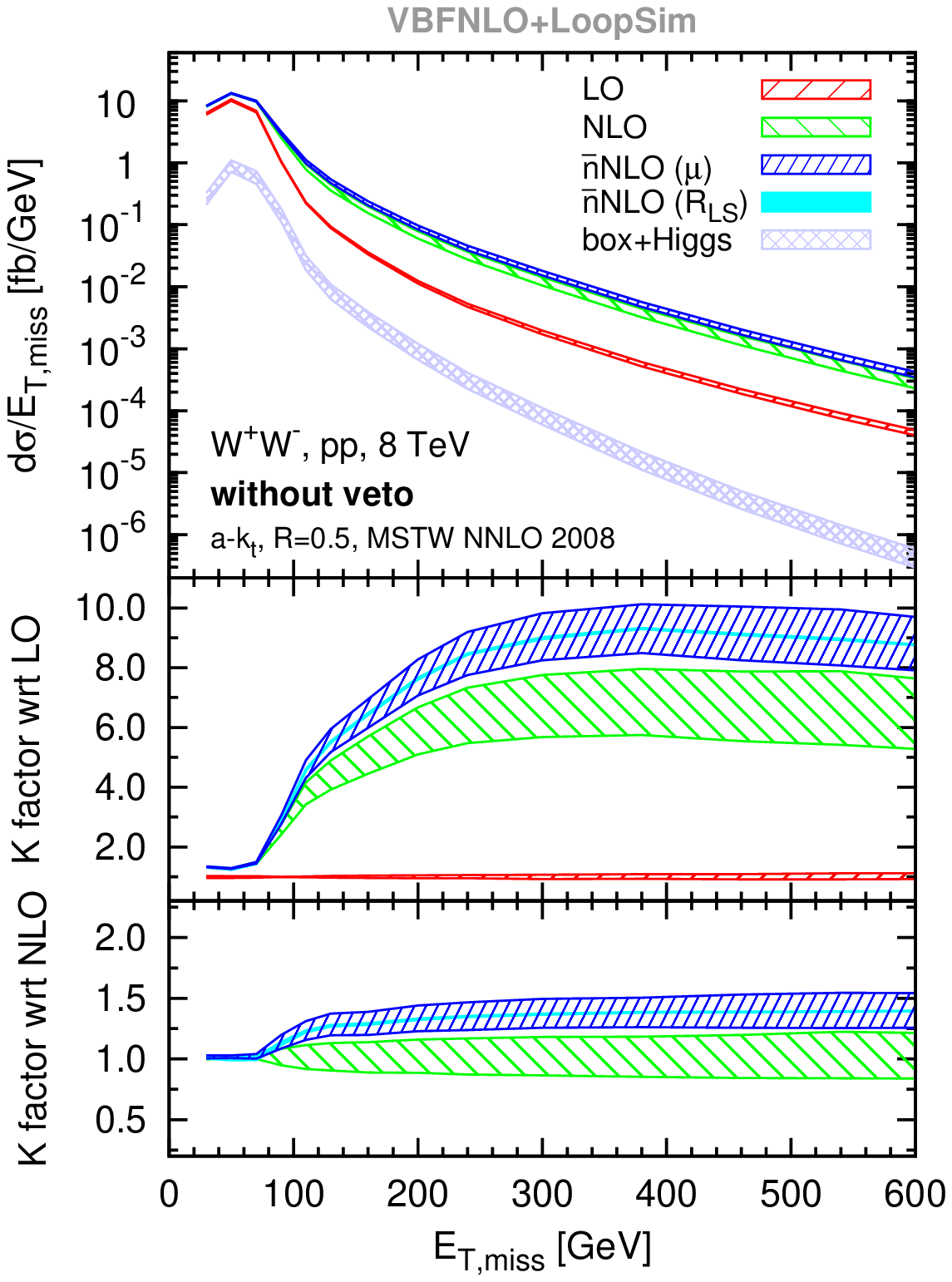}
  \hfill 
  \includegraphics[width=0.49\columnwidth]{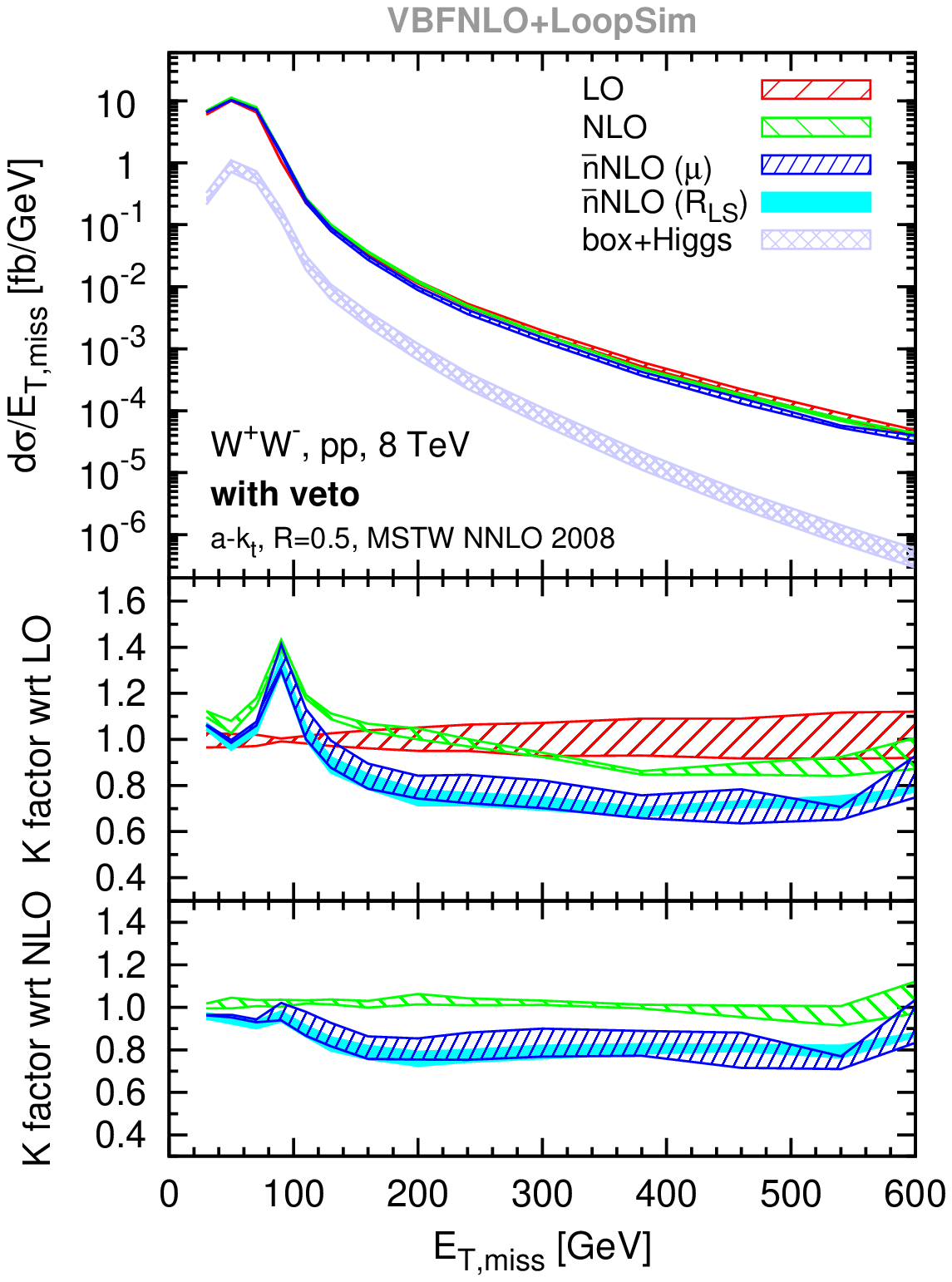}
  \caption{
  Differential cross sections and K-factors for the missing transverse energy 
  for the LHC at $\sqrt{s}=8\, \text{TeV}$ without (left) and with
  jet veto (right). 
  Details are as in Fig.~\ref{fig:HT}.
  }
  \label{fig:ETmiss}
\end{figure}

%%%%%%%%%%%%%%%%%

%%%%%%%%%%%%%%%%%
\begin{figure}[t]
  \centering
  \includegraphics[width=0.49\columnwidth]{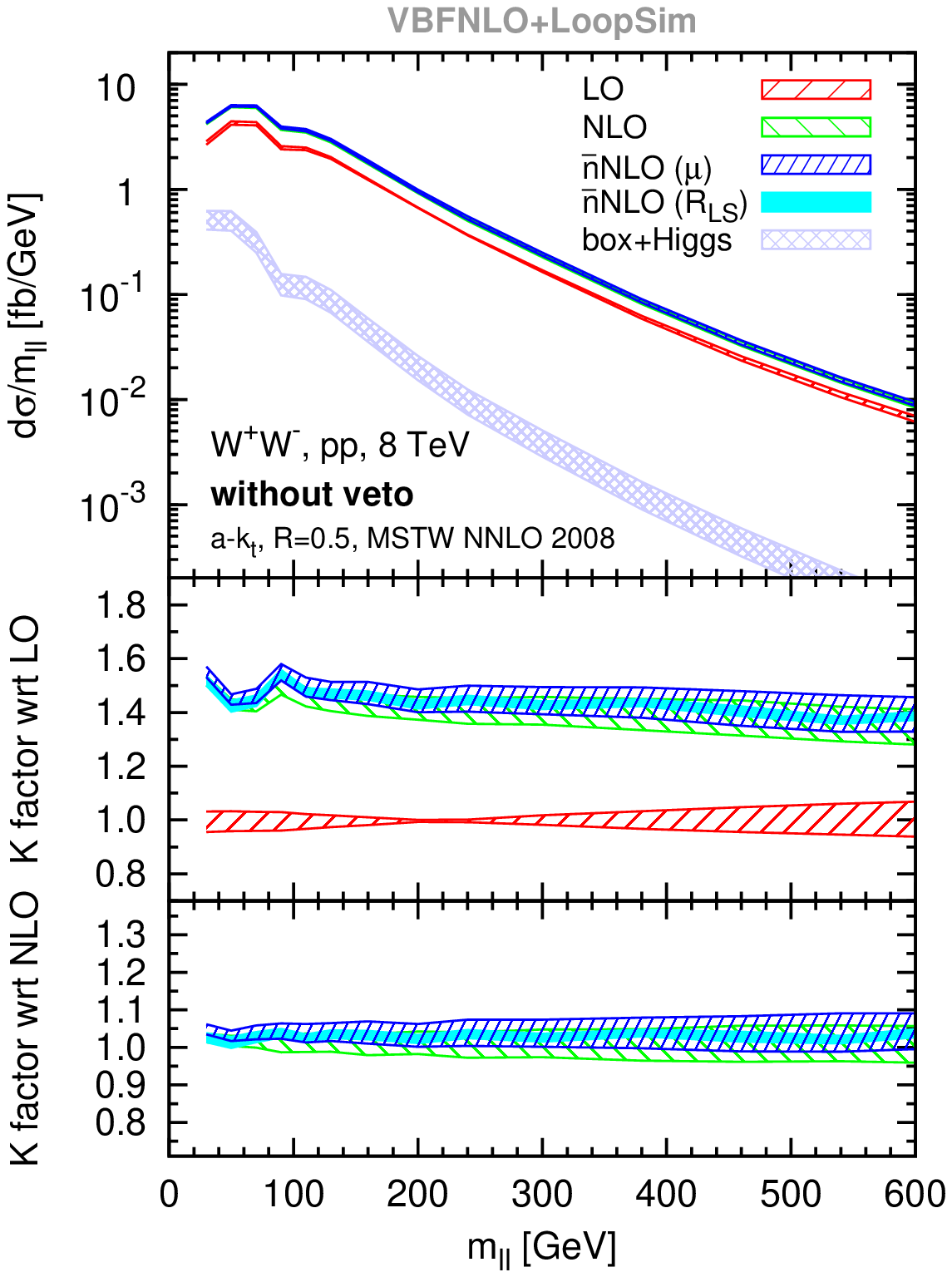}
  \hfill 
  \includegraphics[width=0.49\columnwidth]{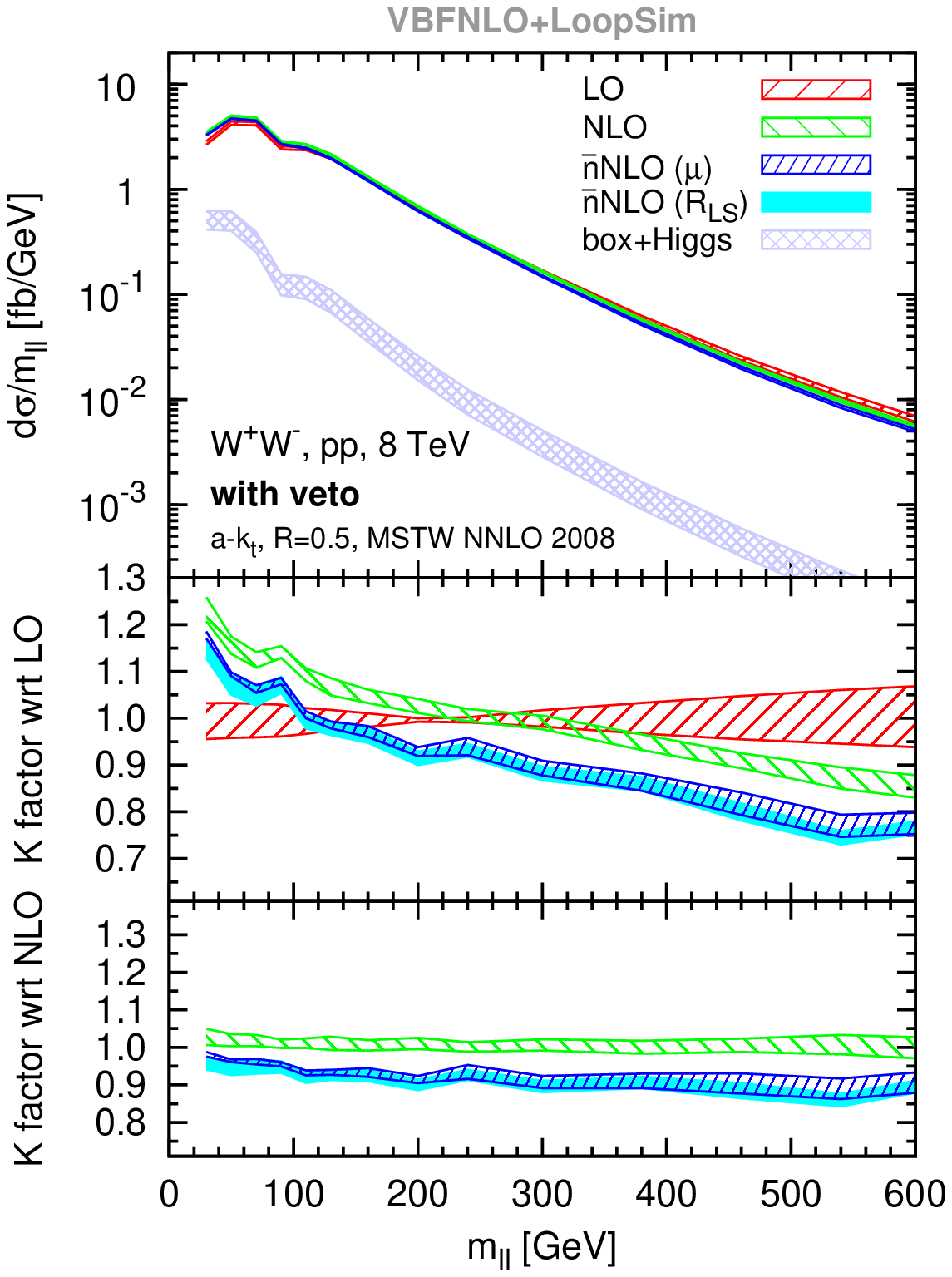}
  \caption{%  
  Differential cross sections and K-factors for the invariant mass of
  the dilepton system for the LHC at $\sqrt{s}=8\, \text{TeV}$ without
  (left) and with jet veto (right). 
  Details are as in Fig.~\ref{fig:HT}.
  }
  \label{fig:mll}
\end{figure}

%%%%%%%%%%%%%%%%%
\begin{figure}[t]
  \centering
  \includegraphics[width=0.49\columnwidth]{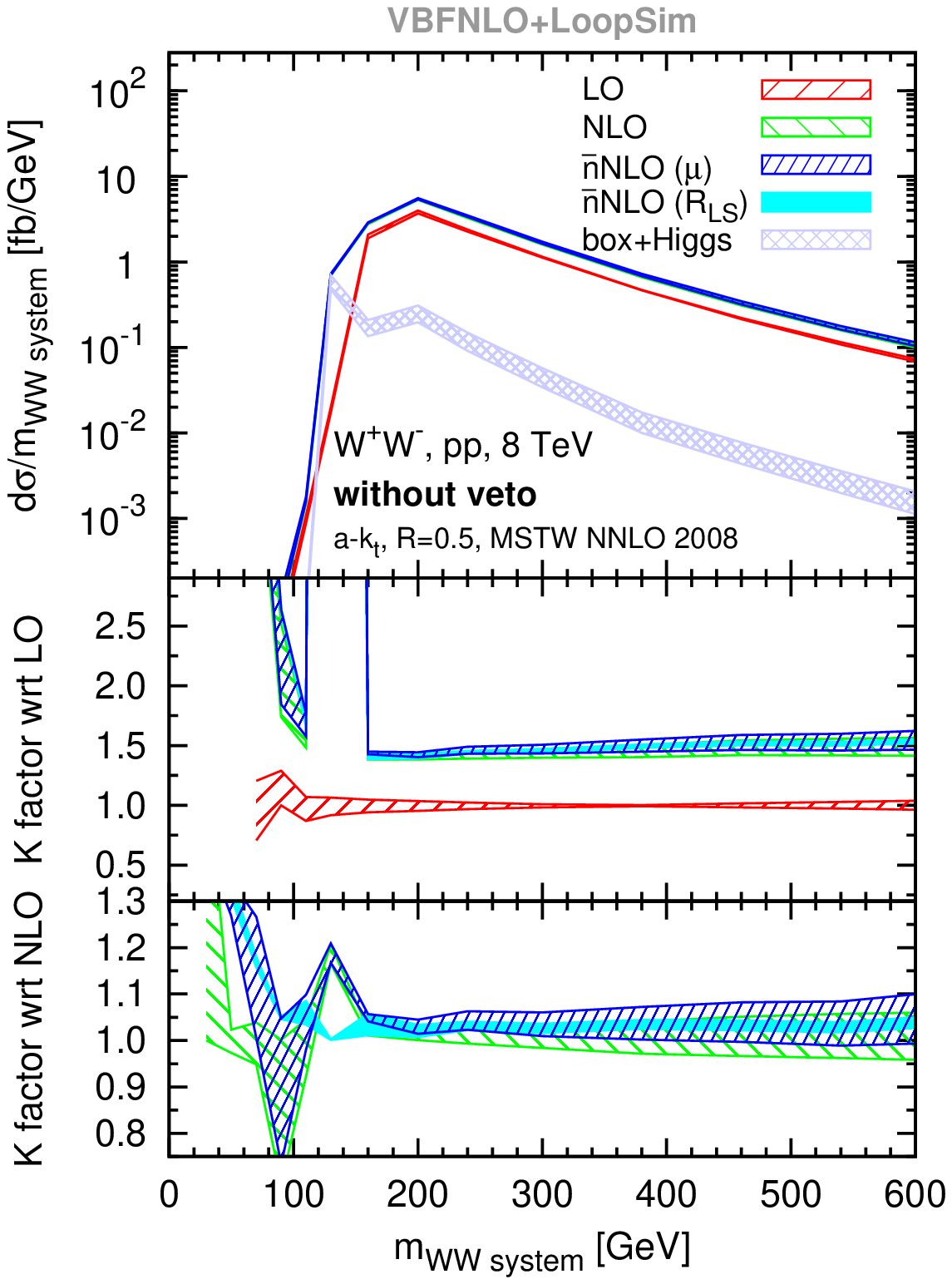}
  \hfill 
  \includegraphics[width=0.49\columnwidth]{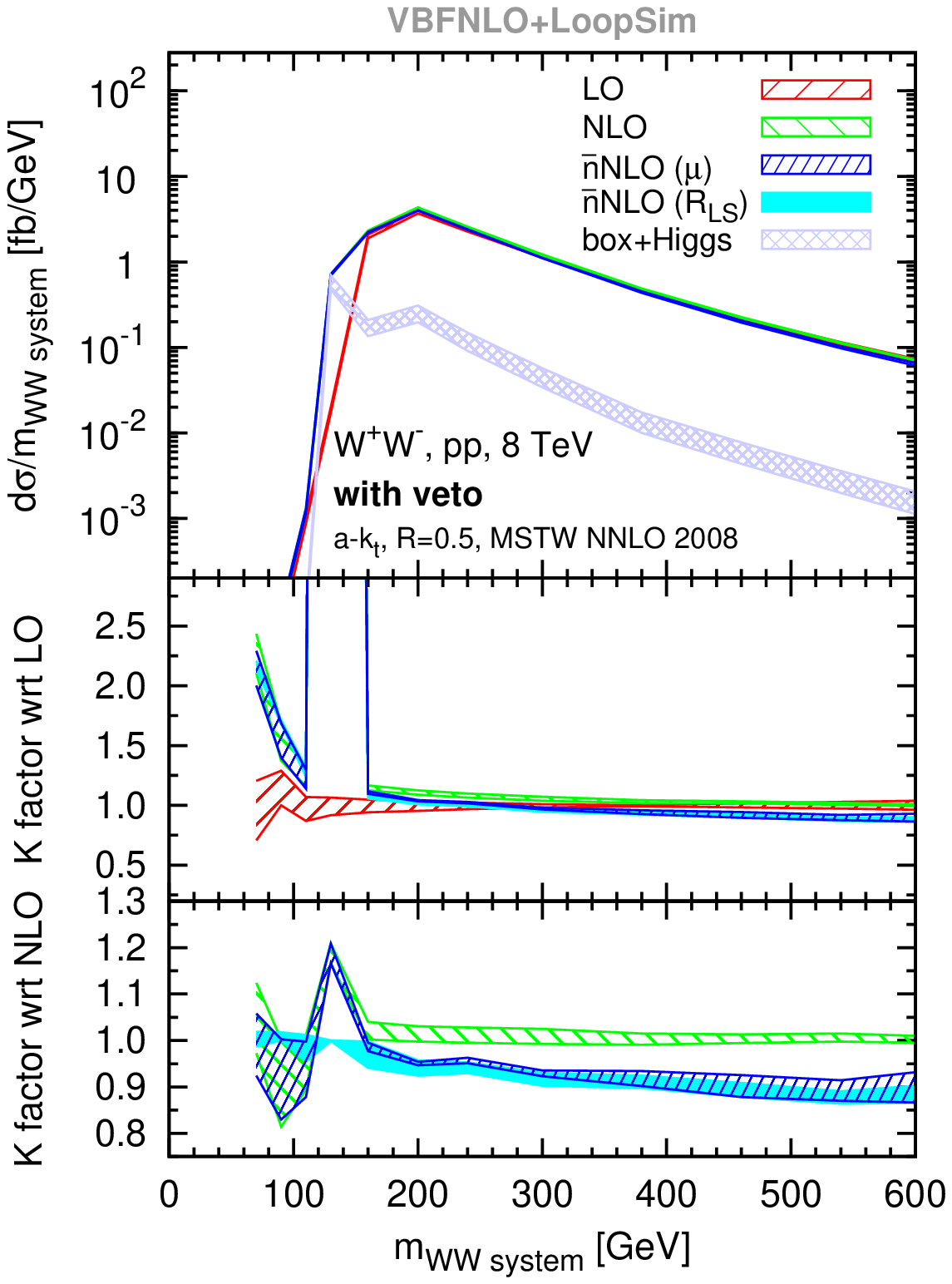}
  \caption{
  Differential cross sections and K-factors for the invariant mass of the diboson system
  for the LHC at $\sqrt{s}=8\, \text{TeV}$ without (left) and with
  jet veto (right). 
  Details are as in Fig.~\ref{fig:HT}.
  }
  \label{fig:mWW}
\end{figure}

In Figs.~\ref{fig:HT}-\ref{fig:mWW} we then present several differential
distributions. On the left-hand side of each figure we show cross sections
without cuts on possible final-state jets, while on the right-hand side,
additionally a jet veto is imposed. The upper panel of each graph shows various
differential cross sections. The NLO and \nNLO curves both include the
contribution from the gluon-fusion diagrams, which is also shown separately with
label ``box+Higgs''. The cyan ``\nNLO ($R_\text{LS}$)'' band shows the
uncertainty from varying the LoopSim parameter, whereas all other bands denote
the change from varying the scale by a factor~$2^{\pm1}$. 
For the results with the gluon-fusion part, \ie~NLO and \nNLO,
the width of the band corresponds to linearly adding errors.
In the middle and the bottom panels we plot the differential K-factor with
respect to LO and NLO, respectively, 
\begin{equation}
K_\text{LO} = \frac{d\sigma / dx}{d\sigma^\text{LO} / dx} \ , \qquad \qquad
K_\text{NLO} = \frac{d\sigma / dx}{d\sigma^\text{NLO} / dx} \ .
\end{equation}
The NLO curve also includes the gluon-fusion contribution.

In Fig.~\ref{fig:HT} we show the effective mass observable $H_T$, defined as
\begin{equation}
H_{T} = \sum  p_{T,\text{jets}} + \sum  p_{T,\ell} + E_{T,\text{miss}}\ ,
\label{eq:HT}
\end{equation}
which is commonly used in new-physics searches. This variable is very sensitive
to additional radiation from further partons and soft or collinear emission of
the W bosons. Once the former pass certain transverse momenta and emission
angles, they yield a significant enhancement of the differential distribution of
$H_T$.  Therefore one expects that without jet veto one observes large K-factors
in the high $H_T$ range. This is indeed seen on the left-hand side of
Fig.~\ref{fig:HT}. For small values of $H_T$, both K-factors are close to unity.
Around a few hundred GeV, however, the K-factors become huge, with values of
$K_\text{LO}$ reaching 12 for NLO and 25 for \nNLO at 1 TeV and still a factor
two for the \nNLO/NLO K-factor. We have also checked that for all
distributions that exhibit a large K-factor, discussed in this paper, the \nLO
result is very close to NLO at $p_T$ above ~200 GeV.

The situation is completely different when we switch on the jet veto. For small
$H_T$ values, the K-factor is again small, but at high $H_T$, the differential
cross section gets significantly reduced as we go from LO to NLO and from NLO to
\nNLO.
This can be easily understood from the definition of $H_T$. An additional
splitting which leads to a further final-state jet will increase $H_T$ while
leaving the partonic center-of-mass energy invariant. Hence, there are
proportionally more events with additional jets at large $H_T$ values. 
By imposing the jet veto, we remove those events and therefore 
kill the huge K-factor seen in the unvetoed plots on the left.
On top of that, the jet veto procedure introduces Sudakov-type logarithms by
forbidding radiation in certain regions of phase space. These logarithms bring
negative corrections to the cross section at high $H_T$. This is precisely what
we observe in the K-factor plots on the right-hand side of Fig.~\ref{fig:HT}.
The NLO/LO and \nNLO/LO K-factors rise a little for small $H_T$, where the
impact of the jet veto is still small. For slightly higher $H_T$, however, the
restriction of additional radiation leads to suppression and eventually fairly
rapid drop of the K-factors.

The cyan solid bands in Fig.~\ref{fig:HT} show the uncertainty due to varying
the $R_\text{LS}$ parameter (related to attributing emission sequence and hard
structure of the events, as explained in Section~\ref{sec:framework}).  For the
vetoed result, this is already smaller than the scale dependence at \nNLO, while
in the unvetoed case this uncertainty is completely negligible.
 
All in all, the \nNLO QCD turns out to bring further negative corrections to
$H_T$ above about 200~GeV for the case with jet veto. One should however be
careful while interpreting these results. 
On one hand, they are potentially subject to further corrections from the
constant term of 2-loop diagrams.%
\footnote{Given that the NLO correction to the
inclusive cross section, which comes predominantly from the constant piece of 1-loop
diagrams, is $\order{30\%}$, a similar correction from the finite term of 2-loop
diagrams could be naively estimated as a square of the 1-loop term, hence it
would amount to $\order{10\%}$.} %
These effects are not accounted for by the
$R_\LS$ uncertainty band.
On the other hand, the non-negligible NLO correction at high $H_T$ 
suggests that the Sudakov logarithms are relevant. Our \nNLO result provides
these type of logarithmic contributions one order higher, hence it supplements
the NLO with an important part of the genuine NNLO correction. To what extent
this negative, Sudakov-type correction is counterbalanced by the finite terms,
can only be checked by performing the full NNLO calculation.
Our approximate \nNLO result for the vetoed case gives already some indications
and an insight into what happens at $\order{\aEW^2\as^2}$. Moreover, as
shown in the following figures, it demonstrates that the small scale
uncertainty of many of the NLO results with jet veto is to a large extent
accidental, as the corresponding uncertainty at \nNLO comes out larger than
that of NLO.

Fig.~\ref{fig:ptlmax} shows another important variable, the transverse momentum
of the hardest lepton, $p_{T,\ell,\max}$, which is extensively used in the
studies of anomalous triple gauge couplings.
Let us first consider the unvetoed results shown on the left-hand side. For
small transverse momenta, where the bulk of the cross section lies, the
differential NLO K-factor is close to the integrated one. 
It then rises when we go to larger $p_{T,\ell,\max}$ values and
finally, above 300 GeV, reaches a plateau at a value of around 2.5. 
The additional \nNLO corrections are very small at the lower boundary, but they
grow to an additional 20\% contribution at large values. Thereby, the NLO and \nNLO
scale variation bands barely overlap, while the uncertainty from $R_\text{LS}$
variation is very small.  
For the actual anomalous gauge coupling searches, an additional jet veto is
imposed to remove events where the whole WW system recoils against jets, which
yield only low sensitivity to aTGC effects. Looking at the vetoed results,
shown on the right-hand side of Fig.~\ref{fig:ptlmax}, we observe that, at large
transverse momenta, the NLO cross section, including the gluon-fusion
contribution, is about 15\% smaller than LO. The \nNLO correction brings an
additional reduction compared to NLO by roughly the same amount. 
Moreover, the scale uncertainty at \nNLO is bigger than at NLO, indicating that
the seemingly small error of the NLO result is largely accidental.
Also here the uncertainty due to varying $R_\text{LS}$ is smaller than the scale
uncertainty.  Hence, the additional contribution from the NLO calculation of WWj
leads to a further reduction of the cross section.

Another interesting observable is the missing transverse energy, $E_{T,
\text{miss}}$, which, again, plays an important role in new physics
searches. Most of the BSM models contain a stable, weakly-interacting
particle, which manifests itself in the detector as a deficit in transverse
energy.
In the SM backgrounds, like WW, this deficit is generated
by two neutrinos in the final state. In
Fig.~\ref{fig:ETmiss} we present the corresponding distribution of missing
$E_T$.  Expectedly, the behaviour is similar to that of the
lepton's transverse momentum shown in the previous figure. The size of
the corrections, however, is even larger here. For the unvetoed cross sections,
the NLO/LO K-factor reaches up to 7 at 600 GeV and the \nNLO
contribution gives an
additional 30\%. Again, the latter is outside the NLO scale variation bands,
while the $R_\text{LS}$ variation is small. The vetoed results, in
contrast, show a decrease of the cross section at larger missing
transverse energy. Thereby, the scale variation uncertainties from NLO
and \nNLO do not overlap, and the latter is still 
larger than the $R_\text{LS}$ error. 
The peak in the NLO/LO and \nNLO/LO K-factors has the same origins as a similar
peak in the $H_T$ distribution discussed above. The strong growth at low $E_{T,
\text{miss}}$ is just the remnant of the large K-factor from the left plot, in
the region where the impact of jet veto is still limited. Then, around 100 GeV,
large negative logarithmic corrections from the veto take over and the K-factor
starts decreasing. Similarly to the $p_{T,\ell,\max}$ case, the scale
uncertainty of the \nNLO vetoed result is larger than that at NLO due to
accidental cancellations occurring in the latter.

In Fig.~\ref{fig:mll} we show the invariant mass of the dilepton system.
This variable plays a crucial role in separating the signal and the control
regions of the WW background in Higgs measurements, although the
relevant energy range there is smaller than the one shown here.
For the unvetoed results, we observe a significant correction similar to the
integrated one when going to NLO, which is basically constant in
$m_{\ell\ell}$. The additional \nNLO corrections, on the other hand, are small
and well within the scale variation uncertainties.
This happens because the distribution from Fig.~\ref{fig:mll} in the high
$m_{\ell\ell}$ region, where configurations with back-to-back leptons dominate,
is not particularly sensitive to new topologies, and the finite terms from the
two-loop diagrams, which are missing in the \nNLO result, are of larger relative
importance for this observable.
The dip in the cross section around the $Z$ boson mass is caused by the cut on
this variable in the same-flavour case.
The results with jet veto show a different behaviour. The NLO/LO K-factor
clearly exhibits a non-constant shape, with positive corrections of about 15\%
at the lower boundary, which gradually falls to negative corrections by over
20\% at the upper end of the shown range. When going to \nNLO, we always observe
a negative correction, which gradually increases in the plotted
range. Thereby, only at very small values the scale uncertainty bands of NLO and
\nNLO overlap, while at larger values the corrections are clearly stronger. The
uncertainty from the $R_\text{LS}$ variation is in the same range as the \nNLO
scale error.

Finally, in Fig.~\ref{fig:mWW} the invariant mass of the WW system is
plotted and shows several interesting features.
At 126 GeV, in the gluon-fusion curve, and hence in the NLO and \nNLO
ones, which include this contribution, the peak from the s-channel Higgs
boson is clearly visible. Then, these cross sections drop again, before,
at twice the W mass, continuum production opens up, which is also
present in the LO curve.
Looking at the K-factors, in the unvetoed case, we see significant
effects from NLO QCD corrections and the loop-induced gluon-fusion
contribution. Beyond that, the additional \nNLO contributions are well
covered by the NLO scale variation for similar reasons as in the $m_{\ell\ell}$
case discussed above.
The distributions with jet veto, depicted in Fig.~\ref{fig:mWW} (right), show
two distinct regions.
Roughly below the WW threshold, we observe the
expected large effects for the K-factor with respect to LO due to the 
gluon-fusion contribution, while the \nNLO effects are small. Above this
value, the NLO corrections become gradually smaller and in the
high-energy range are well within the LO scale variation. The \nNLO part,
on the other hand, gives a gradually growing negative contribution,  
raising up to about 10\% of NLO, which is greater than the bands given by scale
uncertainties and the band due the $R_\text{LS}$ variation.

%-----------------------------------------------------------------------------
\section{Conclusions}
\label{sec:concl}

In this article, we have considered WW diboson production beyond NLO
with W bosons decaying leptonically. This process is an important
background for many new-physics searches as well as SM processes like
the Higgs boson measurements. Furthermore, it is also important as a
signal process for measuring anomalous triple gauge couplings. The NLO
QCD calculations of WW and WW+jet production, as well as the
loop-induced gluon-fusion contribution, implemented in VBFNLO, have been
merged using the LoopSim method to obtain approximate NNLO results for WW
production. The cuts followed closely those used by the two LHC experiments. 

For observables which are sensitive to QCD radiation, like $H_T$ or the
transverse momentum of the leading lepton, we find large additional
corrections beyond NLO. These are typically outside the NLO error bands
given by a scale variation of a factor $\frac12$ and 2. The 
invariant-mass distributions of the dilepton and diboson system, on the
other hand, do not receive significant \nNLO corrections.

Once we impose a veto on jets, as is typically done in the experimental
analyses, we observe further significant negative corrections beyond the NLO
prediction in the high-energy range above approximately 150 GeV, which can be
explained by the appearance of large Sudakov logarithms. 
Their size is larger than the error estimate given by a scale variation
of the NLO cross section and the scale uncertainty of the \nNLO result itself is
larger than that of NLO, which points to an accidental cancellation occurring in the
latter.
The \nNLO vetoed distributions are also potentially subject to non-negligible
corrections from the constant term of the 2-loop diagrams.
Uncertainties due to the LoopSim method, which are estimated by varying
the $R_\text{LS}$ parameter, are always smaller than the remaining scale
uncertainties.
Therefore, we conclude that the QCD corrections to WW production beyond NLO play
an important role for a number of experimentally relevant observables and should
be taken into account in the analyses that use WW theory results as an input.
Until a full NNLO calculation of WW production is available as Monte Carlo
program, our method allows for an inclusion of the dominant part of these
effects. 

The additions to the VBFNLO program used in this article are available
from the authors on request and will be part of a future release.
The LoopSim library, together with the Les Houches Event interface, is publicly
available at \url{https://loopsim.hepforge.org}.

%%%%%%%%%%%%%%%%%%%%%%%%%%
\section*{Acknowledgments} 
We thank Gavin Salam for numerous valuable discussions during all stages of this
work and for useful comments on the manuscript.
FC is funded by a Marie Curie fellowship
(PIEF-GA-2011-298960) and partially by MINECO (FPA2011-23596) and by
LHCPhenonet (PITN-GA-2010-264564).
MR acknowledges partial support by the Deutsche Forschungsgemeinschaft
via the Sonderforschungsbereich/Transregio SFB/TR-9 ``Computational
Particle Physics'' and the Initiative and Networking Fund of the
Helmholtz Association, contract HA-101 (``Physics at the Terascale'').

%%%%%%%%%%%%%%%%%%%%%%%%%%%%%%%%%%%%%%%%%%%%%%%%%%%%%%%%%%%%%%

\end{document}